\documentclass{PoS}

\newcommand{\ovl}[1]{\overline{#1}}

\newcommand{\p}{\partial}

\newcommand{\pslash}{p\kern-1ex /}
\newcommand{\lslash}{l\kern-1ex /}
\newcommand{\kslash}{k\kern-1ex /}
\newcommand{\dslash}{\p\kern-1.2ex /}
\newcommand{\Dslash}{{\cal D}\kern-1.5ex /}
\newcommand{\Aslash}{A\kern-1.2ex /}

\newcommand{\vev}[1]{\left\langle #1 \right\rangle}

\def\tfrac#1#2{{\textstyle\frac{#1}{#2}}}
\title{Chiral properties of light mesons with $N_f=2+1$ overlap fermions}

\ShortTitle{Chiral properties of light mesons with $N_f=2+1$ overlap fermions}

\author{
  JLQCD and TWQCD collaborations:
  \speaker{J.~Noaki}$^{,a}$\thanks{E-mail: noaki@post.kek.jp},
  S.~Aoki${\,}^b$, 
  T.W.~Chiu${\,}^c$,
  H.~Fukaya${\,}^d$,
  S.~Hashimoto${\,}^{a,e}$,
  T.H.~Hsieh${\,}^f$,
  T.~Kaneko${\,}^{a,e}$, 
  H.~Matsufuru${\,}^a$,
  T.~Onogi${\,}^g$,
  E.~Shintani${\,}^g$ and
  N.~Yamada${\,}^{a,e}$
  \vspace*{2mm}
  \\
  \llap{$^a$}
  High Energy Accelerator Research Organization (KEK),
  Tsukuba 305-0801, Japan
  \\
  \llap{$^b$}
  Graduate School of Pure and Applied Sciences,
  University of Tsukuba, Tsukuba 305-8571, Japan
  \\
  \llap{$^c$}
  Physics Department, Center for Theoretical Sciences,
  and Center for Quantum Science and Engineering,
  National Taiwan University, Taipei 10617, Taiwan
  \\
  \llap{$^d$}
  Department of Physics, Nagoya University, 
  Nagoya 464-8602, Japan
  \\
  \llap{$^e$}
  School of High Energy Accelerator Science,
  the Graduate University for Advanced Studies (Sokendai),
  Tsukuba 305-0801, Japan
  \\
  \llap{$^f$}
  Research Center for Applied Sciences, Academia Sinica,
  Taipei~115, Taiwan
  \\
  \llap{$^g$}
  Department of Physics, Osaka University
  Toyonaka, Osaka 560-0043, Japan
}
        
\abstract{ We present an update of the light meson spectrum with
 $N_f$=2+1 overlap fermions on a $16^3\times 48$ lattice
 at five different up and down quark masses and two strange quark masses.
 Based on our experience with the previous simulation with $N_f=2$, we 
 carry out the chiral extrapolation with the prediction of the chiral 
 perturbation theory 
 at the next-to-next-to leading order. 
 We also check the consistency of our analysis by using alternative 
 chiral extrapolation with a reduced theory 
 in which the strange quark mass is integrated out.
}
  
\FullConference{The XXVII International Symposium on Lattice Field Theory\\
		 July 26-31, 2009\\
		 Peking University, Beijing, China}

\begin{document}

\section{Introduction}

The lattice simulation with the overlap fermions~\cite{Neuberger1998}
keeps chiral symmetry for any number of dynamical flavors. 
It implies that the chiral properties of the numerical data 
are described by the continuum chiral perturbation theory (ChPT).
It therefore makes sense to test the consistency between ChPT and QCD.
With this motivation, we performed numerical simulation using
overlap fermion action with two dynamical flavors~\cite{Nf2_spectrum}.
In particular, we carried out the calculation of light meson spectrum.
We fitted the data to the next-to-leading order (NLO) formulae with different 
expansion parameters $m_q$, $m_\pi^2$ and $\xi\equiv m_\pi^2/(4\pi f_\pi)^2$,
all of which should give equivalent prescription at NLO.
As a result, we found that the fit curves start to 
deviate around the scale of kaon mass $m_K$. 
Another important observation was that only
the $\xi$-fit reasonably describes the data beyond the fitted region.
We also demonstrated that next-to-next-to leading (NNLO) effect is 
necessary to describe the data around and beyond $m_K$.

Based on these findings, in this article, we present an extension of 
this study to $N_f=2+1$. 
Using overlap fermion action with $2+1$ dynamical flavors,
we calculate $m_\pi^2/m_{ud}$, $m_K^2/m_{sd}$, $f_\pi$ and $f_K$, where 
$m_{ud}$ is the u-d degenerate mass and $m_{sd}= \tfrac{1}{2}(m_s+m_d)$.
Since these results depend on strange quark mass, the chiral
extrapolation should be performed with a fit ansatz valid beyond
the scale of kaon mass. The only possibility is the fit with the
NNLO ChPT effect taken into account.

After explaining how we obtain the data points briefly in 
Section~\ref{data_pts}, we present in Section ~\ref{SU3_ChPT} the 
chiral extrapolation with NNLO ChPT formulae to treat both pion and kaon 
sectors on an equal footing. Preliminary results of physical values are 
also given in this section. As a consistency check of our analysis, 
in Section~\ref{SU2_ChPT}, we perform the chiral extrapolation with a 
reduced theory in which the strange quark is integrated out. 
Section~\ref{summary} contains a brief discussion about what to be done 
to get to final results on this work. 

\section{Data points}\label{data_pts}

We refer~\cite{MatsufuruProc} for the details of 
the generation of gauge configurations. 
We generate 2,500 trajectories on a $16^3\times 48$ lattice for 
ten combinations of up-down and strange sea quark 
masses, {\it i.e.} five $m_{ud}$'s times two $m_s$'s. 

We calculate 80 pairs of the lowest-lying 
eigenmodes on each gauge configuration and store them on the disks. 
These eigenmodes are used to construct the low-mode contribution to 
the quark propagators. The higher-mode contribution is obtained by 
conventional CG calculation with significantly smaller amount of
machine time than the full CG calculation. Those eigenmodes are also used to 
replace the lower-mode contribution in the meson correlation functions
by that averaged over the source location 
(low-mode averaging)~\cite{DeGrand2004,Giusti2004}. 
We extract meson masses from the exponential decay of the time-separated 
correlation function of pseudo-scalar operator $\vev{P(t)P(0)}$.  
The decay constant, which is defined by the matrix element of
the axial-current operator $A_\mu$, is obtained simultaneously 
using the PCAC relation $\partial_\mu A_\mu = 2m_qP$.

Throughout the Monte Carlo updates, the global topological charge of 
the gauge configurations is fixed to zero. 
This is necessary to avoid discontinuous change of the Dirac eigenvalue,
which is numerically too-expensive.  
The artifact due to fixing the topology is understood as 
a finite size effect~\cite{FSE_Chit}
in addition to the conventional one.
For the physical size of our lattice $L\approx 1.7$ fm, the finite size 
effect could be sizable. We calculate both kinds of finite size effect 
from the analytic formulae based on ChPT~\cite{Colangelo2005,Brower2003}.
In particular, for the effect of the fixed topology, we make use of 
the numerical data of the topological susceptibility
determined on the same lattice configurations~\cite{JLQCD_chit}.

In order to obtain the physical quark mass, we need to renormalize 
bare quark mass on the lattice as $m_q^{\rm (ren)} = Z_mm_q^{\rm (bare)}$.
We obtain the renormalization factor $Z_m$ 
by calculating scalar and pseudo-scalar vertex functions in the momentum 
space in the Landau gauge and applying the RI/MOM scheme~\cite{Martinelli1995}.
In extracting $Z_m$ from the vertex functions,
we control the contamination from the spontaneous chiral symmetry breaking
by using the the low-mode contribution to the chiral 
condensate~\cite{NPR_JLQCD}. Using the perturbative 
matching factor known to 4-loop level and the extrapolation to the 
chiral limit, {\it i.e.} $m_{ud}= m_s=0$, we obtain the result 
$Z_m^{\ovl{\rm MS}}(2\ {\rm GeV}) = 0.806(27)$.

In the rest of this article, it is understood that all data points are 
corrected by the finite size effects and quark masses are renormalized.

\section{Fit to NNLO $SU(3)$ ChPT}\label{SU3_ChPT}

\begin{figure}[t]
 \begin{center}
  \includegraphics[width=7.4cm,clip]{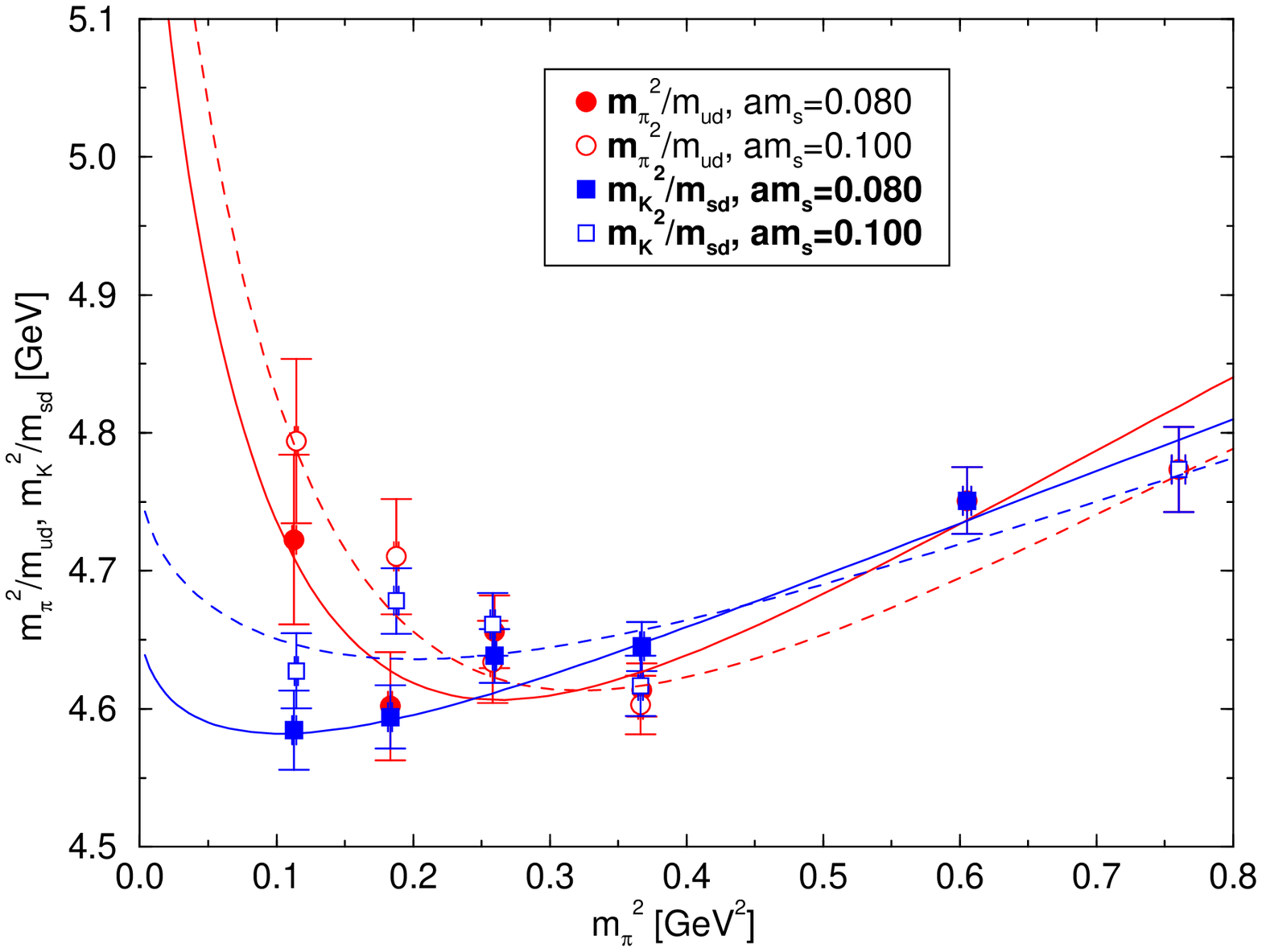}
  \includegraphics[width=7.6cm,clip]{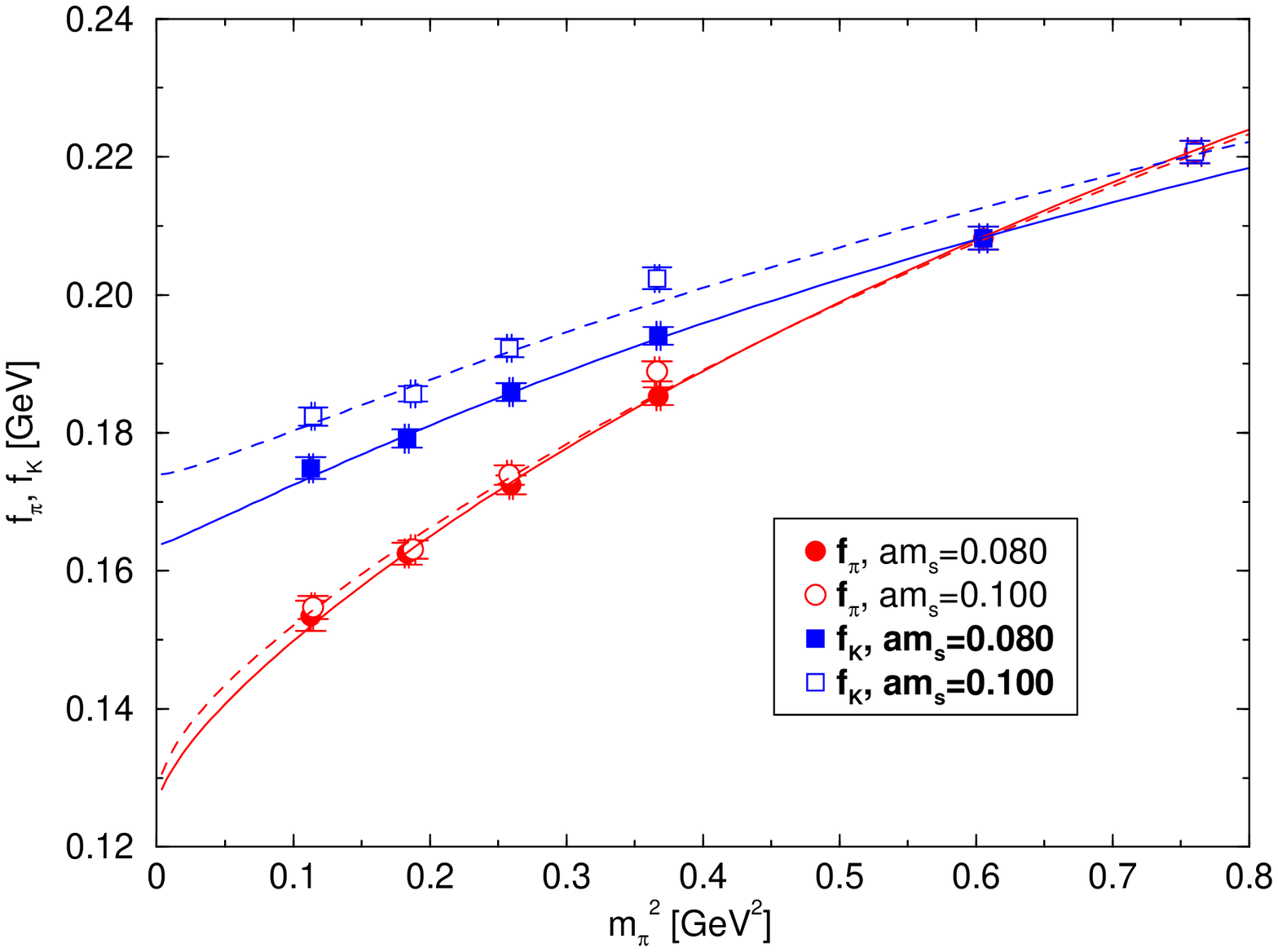}
  \caption{Chiral extrapolation using NNLO full $SU(3)$ ChPT formulae. 
  The left panel contains $m_\pi^2/m_{ud}$ (red circles) and $m_K^2/m_{sd}$ 
  (blue squares) while the right contains $f_\pi$ (red circles)
  and $f_K$ (blue squares). Filled (open) symbols and solid (dashed)
  curves indicate the result with the lighter (heavier) strange quark mass. }
  \label{Nf2+1chpt}
 \end{center}
\end{figure}

Since we found in the two-flavor calculation that the NNLO ChPT formulae
can nicely fit our data even in the kaon mass region 
if one uses the $\xi$-expansion,
we apply the same strategy for our $2+1$-flavor analysis.
As functions of $\xi_\pi=2m_\pi^2/(4\pi f_\pi)^2$ and 
$\xi_K=2m_K^2/(4\pi f_\pi)^2$, predictions from the $SU(3)$ ChPT are 
expressed as 
\begin{eqnarray}
 m_\pi^2/m_{ud} &=& 2B_0
  \left[1+ M^\pi(\xi_\pi, \xi_K; L_4^r,L_5^r,L_6^r,L_8^r)\right] 
  +\alpha_1^\pi\cdot\xi_\pi^2 +\alpha_2^\pi\cdot\xi_\pi\xi_K 
  +\alpha_3^\pi\cdot\xi_K^2, 
  \label{mp2rSU3}\\
 m_K^2/m_{sd} &=& 2B_0
 \left[1+ M^K(\xi_\pi, \xi_K; L_4^r,L_5^r,L_6^r,L_8^r)\right] 
 +\alpha_1^K\cdot\xi_\pi(\xi_\pi -\xi_K) +\alpha_2^K\cdot\xi_K(\xi_K -\xi_\pi),
 \label{mk2rSU3}\\
 f_\pi &=& f_0\left[1+ F^\pi(\xi_\pi, \xi_K; L_4^r, L_5^r)\right]
 +\beta_1^\pi\cdot\xi_\pi^2 +\beta_2^\pi\cdot\xi_\pi\xi_K 
 +\beta_3^\pi\cdot\xi_K^2,
 \label{fpiSU3}\\
 f_K &=& f_0\left[1+ F^K(\xi_\pi, \xi_K; L_4^r, L_5^r)\right]
 +\beta_1^K\cdot\xi_\pi(\xi_\pi -\xi_K) +\beta_2^K\cdot\xi_K(\xi_K -\xi_\pi),
 \label{fKSU3}
\end{eqnarray}
where $m_{sd} = \tfrac{1}{2}(m_s +m_{ud})$.
When we write the chiral Lagrangean as ${\cal L}_\chi = {\cal L}_2
+{\cal L}_4 +{\cal L}_6+\cdots$ with ${\cal L}_n$ indicating the
contribution of ${\cal O}(p^n)$. $\alpha^{\pi,K}_i$ and
$\beta^{\pi,K}_i$ are unknown parameters for the tree-level contribution 
from ${\cal L}_6$. 
Expressions for the functions $M^\pi$, $M^K$, $F^\pi$ and $F^K$ are 
too involved to present here~\cite{Amorosetal2000}. We note that 
each function contains NLO contributions from the one-loop effect 
of ${\cal L}_2$ and tree-level effect of ${\cal L}_4$ and NNLO ones
from the two-loop effect of ${\cal L}_2$ and the one-loop effect of
${\cal L}_4$. These contributions are accompanied by LECs for NLO ChPT,
{\it i.e.} $L_1^r$--$L_8^r$. However, $L_1^r, L_2^r, L_3^r$ and $L_7^r$ 
appear only in the NNLO terms and 
cannot be determined precisely.  We introduce values
$L_1^r =(0.43\pm 0.12)\cdot 10^{-3}$, 
$L_2^r =(0.73\pm 0.12)\cdot 10^{-3}$, 
$L_3^r =(-2.53\pm 0.37)\cdot 10^{-3}$ and 
$L_7^r =(-0.31\pm 0.14)\cdot 10^{-3}$ (defined at $\mu=770$ MeV) 
from a phenomenological estimate~\cite{Amorosetal2001} and determine
others $L_4^r, L_5^r, L_6^r$ and $L_8^r$ by a fit. 
Thus, the chiral extrapolation
with (\ref{mp2rSU3})--(\ref{fKSU3}) contains 16 fit parameters in total.

We fit $m_\pi^2/m_{ud}$, $m_K^2/m_{sd}$, $f_\pi$ and $f_K$
simultaneously taking the correlation within the same sea quark mass 
$(m_{ud}, m_s)$ into account. By using 
all data points, reasonable quality of the fit is obtained with 
$\chi^2/$dof = 2.52.
In this new study, we determine the lattice scale by the result of $f_\pi$
extrapolated to the physical point with the input $f_\pi=130.0$ MeV.
As a result, we obtain $a^{-1} =1.968(39)$ GeV and the pion mass covers
the range of $340\ {\rm MeV} < m_\pi < 870\ {\rm MeV}$.
Figure~\ref{Nf2+1chpt} shows all quantities in question as a 
function of $m_\pi^2$. Different symbols represent the pion data 
($m_\pi^2/m_{ud}$ and $f_\pi$) and the kaon data ($m_K/m_{sd}$ and
$f_K$). Filled (open) symbols represent a
fixed lighter (heavier) strange quark mass, which is accompanied by 
the solid (dashed) curves.

Extrapolating the data to the physical point 
$(\xi_\pi^{\rm (phys)}, \xi_K^{\rm (phys)})$, which is determined with
$m_{\pi}=135.0$ MeV, $m_K=495.0$ MeV and $f_\pi=130.0$ MeV, we obtain 
preliminary results
$m_{ud}^{\ovl{\rm MS}}(2\ {\rm GeV}) = 3.64(12)\ {\rm MeV}$,
$m_s^{\ovl{\rm MS}}(2\ {\rm GeV}) = 104.5(1.8)\ {\rm MeV}$,
$m_s/m_{ud} = 28.71(52)$, $f_K   = 157.3(5.5)\ {\rm MeV}$ and
$f_K/f_\pi = 1.210(12)$,
where the errors are statistical only. 

\section{Fit to the reduced $SU(2)$ ChPT to NLO}\label{SU2_ChPT}

\begin{figure}[t]
 \begin{center}
  \includegraphics[width=7.4cm,clip]{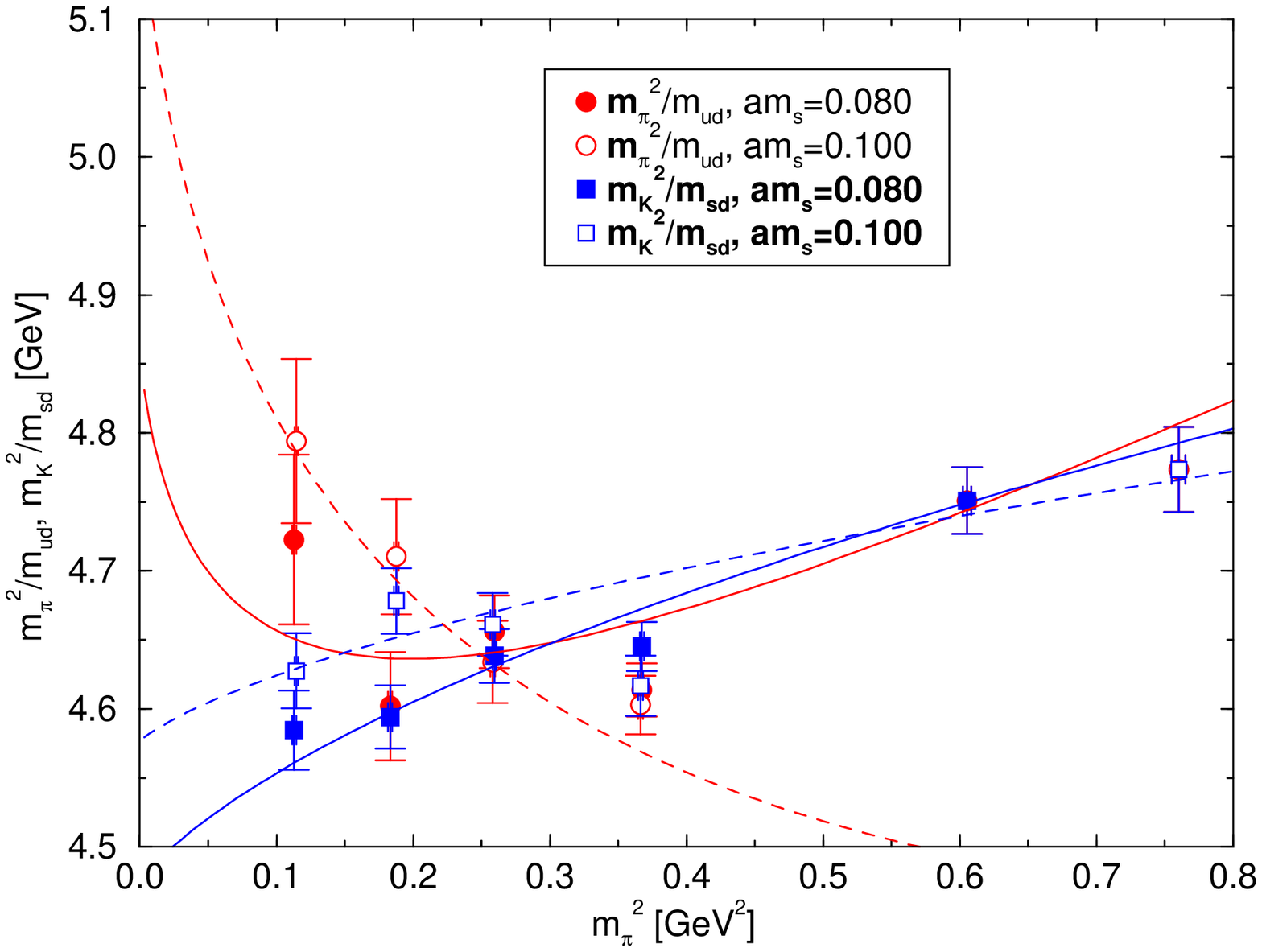}
  \includegraphics[width=7.6cm,clip]{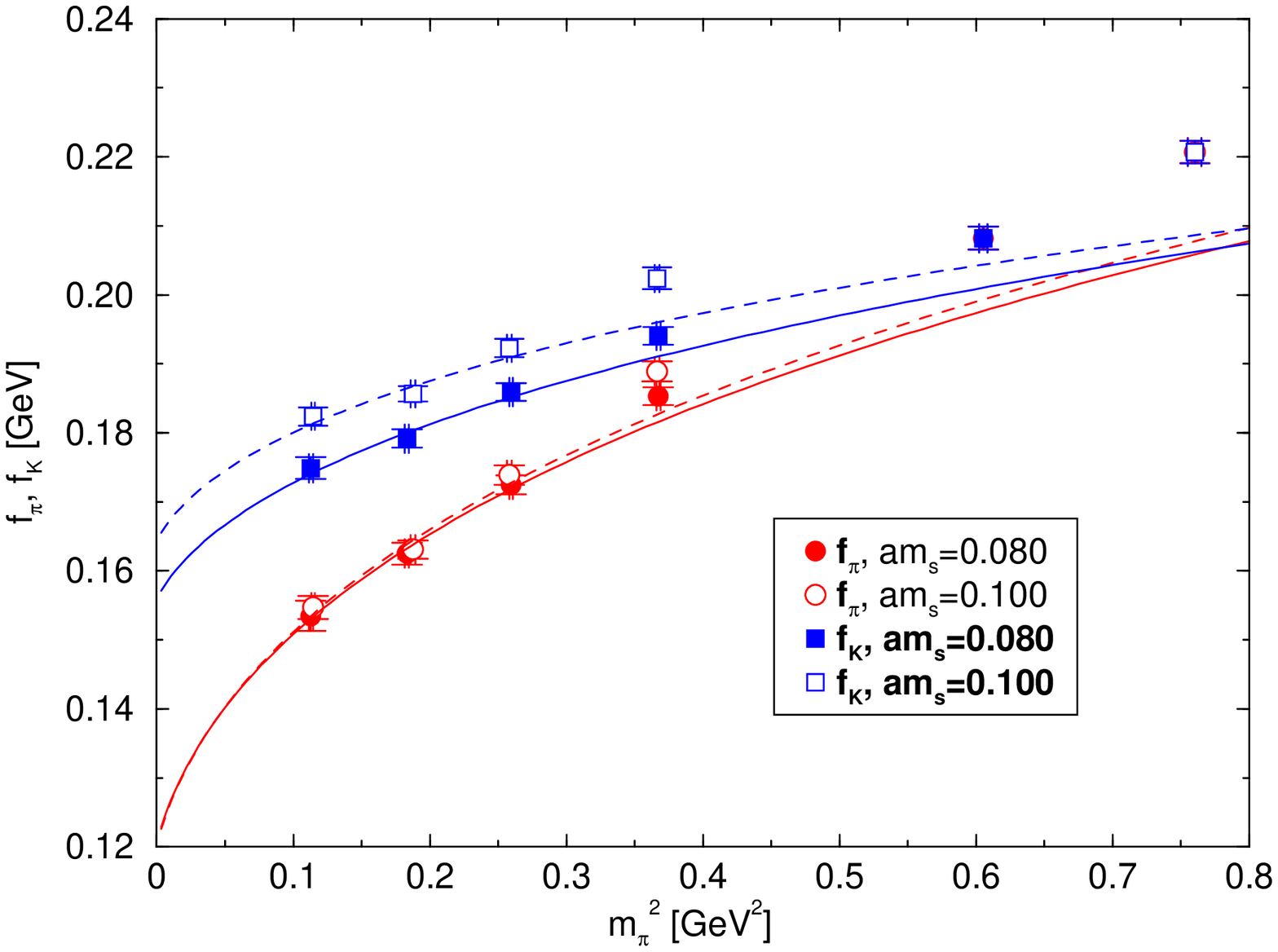}
  \caption{Chiral extrapolation using the prediction from 
  the reduced $SU(2)$ ChPT to NLO. 
  The organization is same as Figure~1. For the fit, the lightest 
  three data points are used for each quantity.}
  \label{Nf3_su2NLO}
 \end{center}
\end{figure}

As a check of the chiral extrapolation with the NNLO ChPT in the
previous section, we also study a different fit ansatz.
It is also possible to carry out the extrapolation to the physical 
point $\xi_\pi^{(\rm phys)}$ by paying 
attention only to the dependence on the up-down quark mass,
or the pion mass. Integrating out the strange quark as a 
static heavy quark in ChPT, one obtains an effective theory which
respects a reduced 
$SU(2)$ symmetry~\cite{Gasseretal2007,RBC_UKQCD_spect,PACSCS_spect}. 
At NLO, the chiral expansion reads
\begin{eqnarray}
 m_\pi^2/m_{ud} &=& 2B\left( 1 +\tfrac{1}{2}\xi_\pi\ln\xi_\pi\right) 
  +c_3\,\xi_\pi,\\
 m_K^2/m_{sd} &=&  2B^{(K)} +c_1^{(K)}\,\xi_\pi,\\
 f_\pi &=& f \left(1 -\xi_\pi\ln\xi_\pi\right) +c_4\, \xi_\pi,\\
 f_K &=& f^{(K)}\left( 1 -\tfrac{3}{8}\xi_\pi\ln\xi_\pi\right) 
 +c_2^{(K)}\xi_\pi,
\end{eqnarray}
where we have LECs $B^{(K)}$, $f^{(K)}$, $c_1^{(K)}$ and 
$c_2^{(K)}$ in addition to the LECs for $SU(2)$ ChPT, {\it i.e.} 
$f$, $B$, $\bar{l}_3$ and $\bar{l}_4$.
In the present case, all LECs depend on strange quark mass.
With the lightest three $m_{ud}$ points, which are in the 
valid region of this framework, {\it i.e.} $m_{ud}\ll m_s$ for 
each fixed value of $m_s$, we carry out the correlated fit for the 
quantities sharing the same mass point $(m_{ud}, m_s)$.
Figure~\ref{Nf3_su2NLO} shows the fit curves obtained in this way.

\begin{figure}[t]
 \begin{center}
  \includegraphics[width=6.8cm,clip]{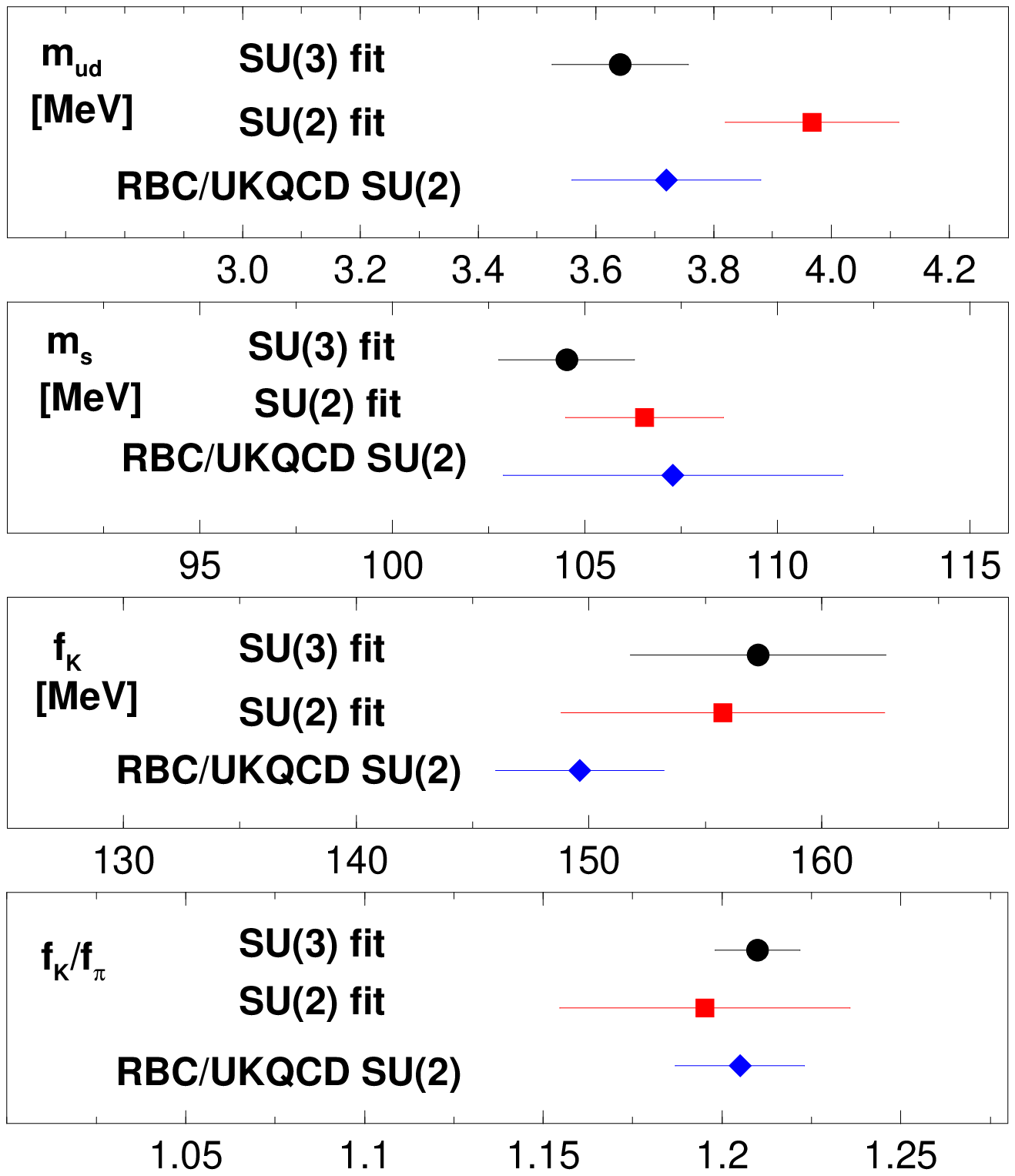}
  \hspace{0.6cm}
  \includegraphics[width=6.8cm,clip]{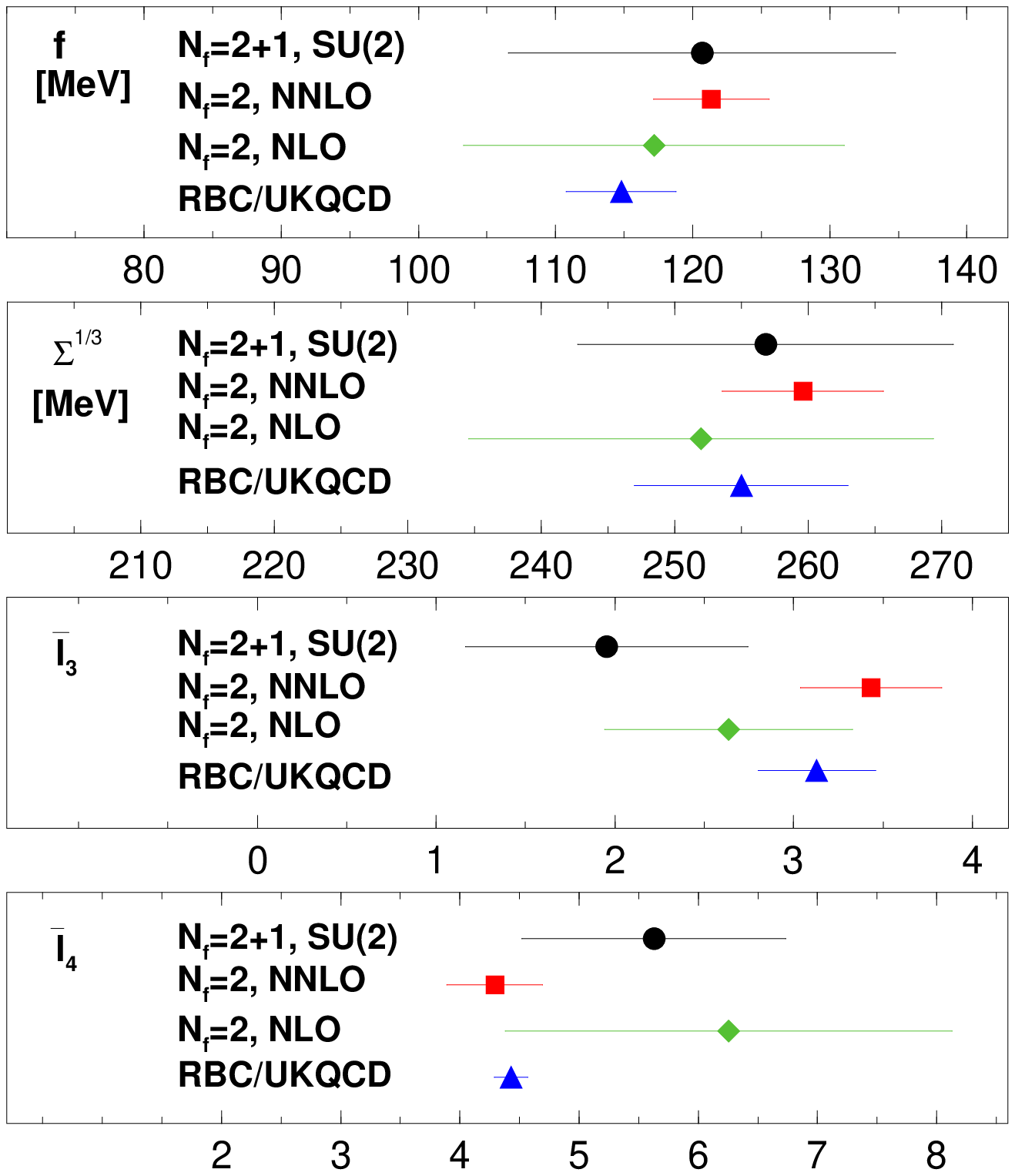}
  \caption{Left: comparison of physical values for full QCD 
  obtained by different chiral extrapolations.
  Right: comparison of the $SU(2)$ LECs between different dynamical 
  flavor $N_f$. All indicated errors are only statistical.}
  \label{su2_vs_su3}
 \end{center}
\end{figure}

The fit results for each fixed $m_s$ are extrapolated to the physical 
strange quark mass $m_s^{\rm (phys)}$, which is determined by solving
$m_K^2/m_s|_{\xi_\pi^{(\rm phys)}} = (495.0 {\rm MeV})^2/m_s$.
In the light panel of Figure~\ref{su2_vs_su3}, we compare physical 
results for $m_{ud}$, $m_s$, $f_K$ and $f_K/f_\pi$ from the full NNLO $SU(3)$
ChPT (circles), and from the NLO reduced $SU(2)$ ChPT (squares from our 
analysis and diamonds from the similar analysis by RBC and UKQCD 
Collaborations~\cite{RBC_UKQCD_spect}). 
The reasonable agreement among different fitting prescriptions
provides a good consistency check of the analysis.

It is also interesting to compare $SU(2)$ LECs obtained in this work
with our previous work with $N_f=2$.
In the right panel of Figure~\ref{su2_vs_su3}, results of $f$,
$\Sigma=Bf^2/2$, $\bar{l}_3$ and $\bar{l}_4$ from the $N_f=2+1$ simulation  
(circles for the present work and triangles for \cite{RBC_UKQCD_spect}) 
are compared with the results obtained in \cite{Nf2_spectrum} 
(squares from the NNLO fit and diamonds from the NLO fit).
The reasonable agreements observed for each quantity
imply that the reduced $SU(2)$ ChPT intermediates full theories with 
$N_f=2$ and $2+1$.

\section{Summary}\label{summary}

We studied the chiral property of ChPT by comparing the analytic 
prediction with the lattice data obtained in the $N_f=2+1$ dynamical 
simulation with the overlap fermions. 
We fitted the data to the ChPT prediction at NNLO for the first time. 
It is the only way to describe both pion and kaon data on an uniform 
basis and to study the convergence property of the $SU(3)$ ChPT.
The validity of the extrapolation to the physical mass point is checked 
with the results from the fit with the reduced $SU(2)$ ChPT. 
However, in order to discuss the convergence property of ChPT as in
the case of $N_f=2$, we need to determine individual LECs of the $SU(3)$
ChPT with good accuracy. 
With the data points obtained for two different strange quark masses, 
we have a limited constraint on the $\xi_K$ dependence hence large errors 
for LECs. 
For example, in (\ref{fpiSU3}) or (\ref{fKSU3}), $f_0$ cannot be 
determined unambiguously unless the contribution of the terms with 
$L_4^r$ and $L_5^r$ are well fixed. 
Moreover, from the phenomenological side, it is advantageous to determine 
LECs from this work because the results can be used as inputs 
in the calculation of different quantities including $B_K$ and $K_{l3}$
form factors. For these motivations, 
we are generating more data points with $m_{ud}=m_s$ to obtain 
the $SU(3)$ LECs with high accuracy.

Another work in progress is the calculation on a larger volume for 
a direct check of finite size effect (FSE). We are generating the data 
for the two lightest $m_{ud}$'s and the lighter $m_s$ on a 
$24^3\times 48$ lattice with the same coupling constant as the 
present work. We are also planning to obtain the lattice scale
from $\Omega$-baryon mass on the larger volume to compare with the
current value from the $f_\pi=130$ MeV input.

\vspace*{6mm}

Numerical simulations are performed on Hitachi SR11000 and
IBM System Blue Gene Solution at High Energy Accelerator Research
Organization (KEK) under a support of its Large Scale
Simulation Program (Nos.~08-05 and~09-05 ). 
The work of HF was supported by the Global COE program of Nagoya
University "QFPU" from JSPS and MEXT of Japan.
This work is supported in part by the Grant-in-Aid of the 
Ministry of Education
(Nos.
19540286,
19740121,
19740160,
20105001,
20105002,
20105003,
20105005,
20340047,
21105508,
21674002)
and the National Science Council of Taiwan
(Nos. NSC96-2112-M-002-020-MY3,
      NSC96-2112-M-001-017-MY3,
      NSC98-2119-M-002-001),
and NTU-CQSE (Nos. 98R0066-65, 98R0066-69).

%%%%%%%%%%%%%%%%%%%%%%%%%%%%%%%%%%%%%%%%%%%%%%%%%%%%%%%%%%%%%%%%%%%%%%
%\newcommand{\J}[4]{{\it #1} {\bf #2} (19#3) #4}
\newcommand{\J}[4]{{#1} {\bf #2} (#3) #4}
\newcommand{\RMP}{Rev.~Mod.~Phys.}
\newcommand{\MPL}{Mod.~Phys.~Lett.}
\newcommand{\IJMP}{Int.~J.~Mod.~Phys.}
\newcommand{\NP}{Nucl.~Phys.}
\newcommand{\NPSup}{Nucl.~Phys.~{\bf B} (Proc.~Suppl.)}
\newcommand{\PL}{Phys.~Lett.}
\newcommand{\PRD}{Phys.~Rev.~D}
\newcommand{\PRL}{Phys.~Rev.~Lett.}
\newcommand{\AP}{Ann.~Phys.}
\newcommand{\CMP}{Commun.~Math.~Phys.}
\newcommand{\CPC}{Comp.~Phys.~Comm.}
\newcommand{\PTP}{Prog. Theor. Phys.}
\newcommand{\Suppl}{Prog. Theor. Phys. Suppl.}
\newcommand{\JHEP}{JHEP}
\newcommand{\PoS}{PoS}
%%%%%%%%%%%%%%%%%%%%%%%%%%%%%%%%%%%%%%%%%%%%%%%%%%%%%%%%%%%%%%%%%%%%%%

\end{document}